\documentclass[conference]{IEEEtran}
\IEEEoverridecommandlockouts
\usepackage{hyperref}
\usepackage{amsmath,amssymb,amsfonts}
\usepackage{algorithmic}
\usepackage{graphicx}
\usepackage{textcomp}
\usepackage{xcolor}

\usepackage{listings}
\usepackage{todonotes}

\begin{document}

\title{Research Data Version Management and Machine-Actionable Reproducibility for HPC}
 
\author{\IEEEauthorblockN{Andreas Knüpfer}
\IEEEauthorblockA{\textit{Center for Advanced Systems Understanding (CASUS)} \\
\textit{Helmholtz-Zentrum Dresden-Rossendorf (HZDR)}\\
Görlitz, Germany \\
\href{mailto:a.knuepfer@hzdr.de}{a.knuepfer@hzdr.de}, \href{https://orcid.org/0000-0003-3591-397X}{ORCID 0000-0003-3591-397X}}
\and
\IEEEauthorblockN{ Timothy J. Callow}
\IEEEauthorblockA{\textit{Center for Advanced Systems Understanding (CASUS)} \\
\textit{Helmholtz-Zentrum Dresden-Rossendorf (HZDR)}\\
Görlitz, Germany \\
\href{https://orcid.org/0000-0002-4878-3521}{ORCID 0000-0002-4878-3521}}
}

\maketitle

\begin{abstract}
We present a solution for research data version control and machine-actionable reproducibility of data processing for High Performance Computing (HPC) environments and the SLURM batch scheduler.
Both aspects are important for research data management and the DataLad tool provides both based on the very prevalent git version control system. However, it is incompatible with HPC batch processing.
The presented extension makes it compatible with HPC batch processing with the SLURM scheduler. It solves the fundamental incompatibility so that multiple jobs can be scheduled concurrently on the same data repository. It also avoids inefficient behavior patterns which may emerge on parallel file systems.
\end{abstract}


\lstset{
language=bash,
basicstyle=\bf\ttfamily,
breakatwhitespace=false,
showspaces=false,
backgroundcolor=\color{lightgray}
}
\newcommand{\shell}[1]{\lstinline{#1}}

\section{Introduction and Motivation: Working with evolving data collections}

The presented solution enables version control for large data collections in HPC environments in a fully machine-actionable reproducible manner.
The poster example is the training process of DNN surrogate models from results of extensive HPC simulations \cite{mala1,mala2}. Assume millions of CPUh are required to produce training data with standard simulation programs over the course of several months. 
The DNN training is equally demanding in terms of HPC computations. It will be done multiple times to check feasibility and for hyperparameter optimization. This will start early with a sub-set of the training data. Later it may be re-generated or fine-tuned as successive version of the DNN model with more training data when this becomes available.
So we have two large, evolving datasets in the process of being completed concurrently. Multiple people might contribute at the same time using multiple HPC clusters. Data will be added in continuously or sporadically. Sometimes data points might be found to be faulty and need to be removed or replaced.

The question \textit{"Which was the precise subset and version of the dataset used for the training process?"} is a challenging one. However, it is most interesting for fellow scientists who want to reproduce results and build on top of it. They would want to know if a given version still contained the faulty data point. Or for the funding agency who demands reproducible research. Or when the investigation board asks for the exact training dataset after that space probe incident.

Research data version control is even more challenging when redundant copies of all intermediate versions are infeasible. Even though this may work for small datasets or for infrequent data publications (see Section~\ref{sect:ref:reproduce}) it is inefficient. For large data collections and fine-grained versioning this is out of the question. Interestingly, data versioning also goes beyond the F.A.I.R. criteria (see Section~\ref{sect:notinfair}).

For research software and software development in general this problem is solved (see Section~\ref{sect:versioning}). The same principles and the same tools have been applied to research software data management successfully (see Section~\ref{sect:versioningfordata}). Yet, this is not applicable for HPC and thus the most computationally and data intensive research. The presented solution will solve this.

\section{Related Work and Background}
\label{sect:related_work}


\subsection{The F.A.I.R. principles without versioning?}
\label{sect:notinfair}

The F.A.I.R. principles are the widely accepted criteria for good research data management and part of many institutional guidelines as well as national and international standards \cite{wilkinson2016fair}.

Reproducibility as part of provenance is a core aspect of the F.A.I.R. principles but versioning of data is missing. Only criterium \textit{"A2. metadata are accessible, even when the data are no longer available"} touches versioning most rudimentarily \cite{wilkinson2016fair}.
So, was versioning overlooked when the "F.A.I.R." criteria were defined? Or was it seen as a mere technical detail as part of data provenance? Or deemed to much to ask back when F.A.I.R. was proposed? The authors assume the latter because access to (some) past versions of datasets would simply be too useful to be left out.

Very few research data management papers address version control for data as a key requirement.
Klump et.al. \cite{klump_2020_3772870} describe the state of the art in research data versioning which are explicitly defined \textit{data publications} as coarse-grained version steps. They comment how it is not as fine-grained as source code versioning and causes redundant storage, still they recommend it as the minimum standard for research data.
Heinrichs et.al. \cite{eval_of_rdm_architectures} define requirements for research data management and include versioning of data and metadata. 
It is not clear if this refers to coarse-grained or fine-grained versioning.
Freund et.al. \cite{RDM_version_control} recently compared a list of solutions including git-LFS and DVC for their data version control capabilities and found that non of them fulfilled all their defined requirements. 
%

\subsection{Version control for source code}
\label{sect:versioning}


\shell{git} is the de-facto software tool for distributed source code version control and collaborative development. It is applicable for all kinds of text-based documents and relies on the ability to track changes via \shell{diff} across revisions instead of storing redundant copies of text files. Therefore, it always keeps \textit{all} past revisions of \textit{all files} in each branch. Standard \shell{git} can handle only small to medium sized binary files if they don't change often. For details and git terminology see the git documentation\footnote{The official git documentation \url{https://git-scm.com/doc}}.

\subsection{Version control tools for research data} 
\label{sect:versioningfordata}


The \shell{git-annex} extension allows managing large files with \shell{git}, without storing the file contents in \shell{git}. It is an extension to the \shell{git} command which adds a new sub-command "\shell{annex}". When working with the usual \shell{git} sub-commands, it intercepts so called \textit{annexed} files (think large binary files but basically any file or file type configured) and handles them in a special way. The main difference is that the file contents is not stored in the git repository, only the file name and some metadata are kept there as part of the version management.
The data itself is kept in distributed key-value storages which can be shared across all clones and all branches of a repository. There are many supported storage protocols for annexed files, among them S3, webdav, git-LFS and rclone%
\footnote{Rclone \url{https://rclone.org/} can access almost all remote storage protocols.}.

\begin{figure}[t]
\centering
\includegraphics[scale=0.2]{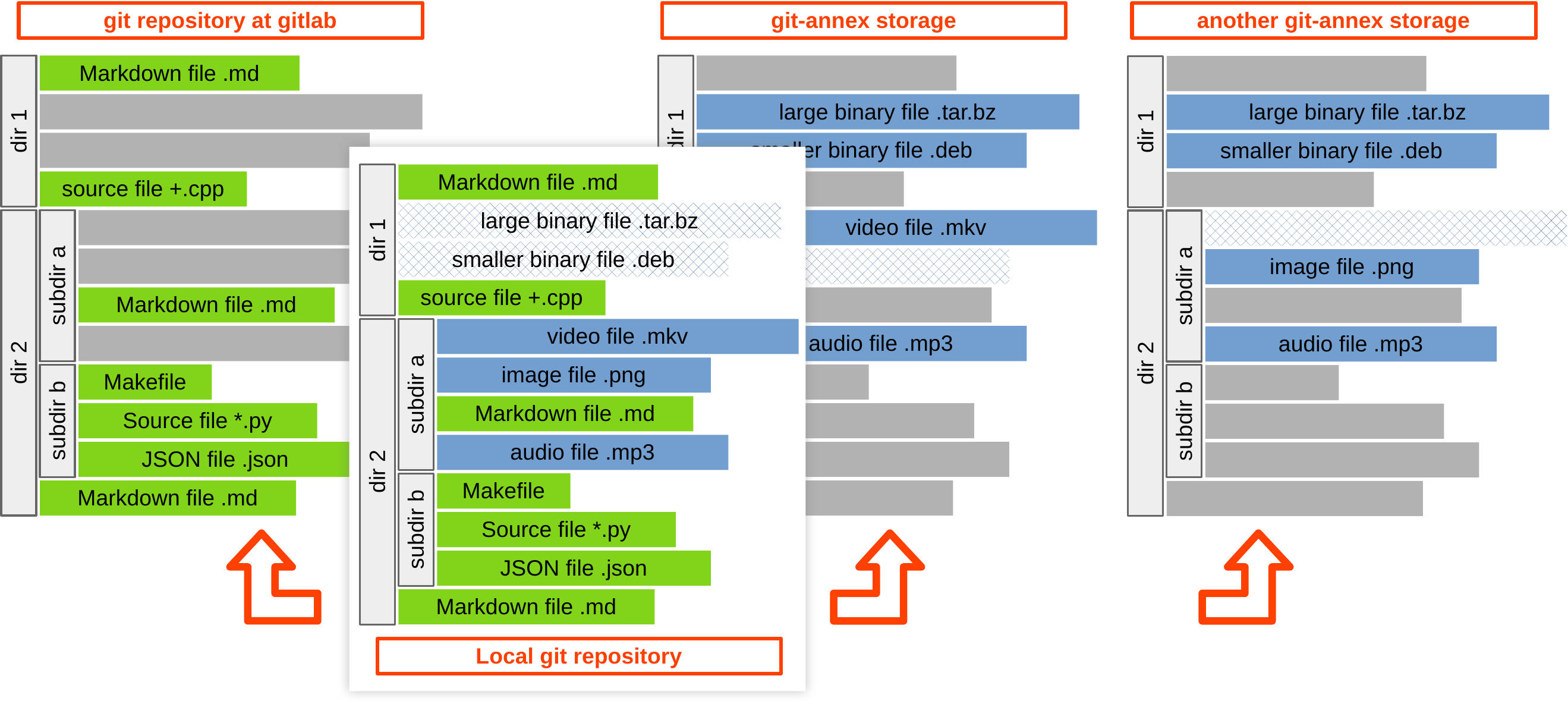}
\caption{Overview of a local git repository with annexed files. It connects to the \textit{remote origin} at gitlab which holds the full set of text files (green).
And it connects to one or multiple annex storages which can hold different subsets of the annexed files (blue). The contents of an annexed file only needs to be stored in one of them, not in all (hatched) -- unlike files in a git repository.}
\label{fig:git-annex-repo}
\end{figure}

\shell{git-annex} abandons the guarantee that all revisions in a branch can be re-created from the local repository alone. It even goes one step further, in that the file content of annexed files is not present by default (not event the most recent version). That means after cloning a DataLad repository the annexed files are known but their content is not present. This is actually an advantage, because one can clone a repository on a laptop and over a standard network connection even though the total size
including all the annexed files
would be many terabytes\footnote{Compare the DataLad superdataset at \url{https://datasets.datalad.org/}}.
The \shell{git annex get} sub-command will get all / some files on demand from one of the annex storages. \shell{git} \shell{annex drop} will remove a local copy again.
\figurename~\ref{fig:git-annex-repo} gives an overview about the repositories and storages.

Git Large File Storage (Git LFS) is another Open Source extension for git designed to manage large files and binary data. Like git-annex it replaces large/binary files in a git repository with lightweight pointers, while the actual file content is stored in a separate location.
Git LFS focuses more on simplicity and (almost) works without additional git sub-commands. Compared to git-annex it lacks some flexibility like multiple annex storages or offline features.
Still, both are very similar and fulfill the same purpose.
The presented approach uses git-annex because it is part of DataLad.


LakeFS is a web front-end for S3 storage which provides versioning for large data sets or "data lakes" and offers "Git-like capabilities"%
\footnote{The LakeFS webpage at \url{https://docs.lakefs.io/project/}}. Compared to git it provides only the basic versioning, branching, and merging functionality.
The LakeFS web front-end includes integrated database querying functionality. There are also a REST API and a command line interface.
LakeFS consists of an Open Source core under Apache 2.0 license and a commercial product on top of it. The authors experienced in the past that the Open Source version is severely limited, however. 
There are a number of further tools which are specific for different research communities \cite{RDM_version_control}.

\subsection{Reproducibility solutions suitable for HPC}
\label{sect:ref:reproduce}

The standard solution for machine-actionable reproducibility in data-driven research are workflow automation systems like AiiDA \cite{huber2020aiida, uhrin2021aiida}, Galaxy \cite{galaxy2024galaxy,gruning2018galaxyreproducibility}, or similar. Different solutions are favored in different fields of science.
All offer multi-stage "workflows" or "processing pipelines" as combination of processing steps from lightweight pre/postprocessing to heavyweight parallel computations.
They manage many instances for many datasets and aim at high-throughput computing (HTC, many independent jobs) but also integrate with HPC.

Workflow managers keep track of input / intermediate / output datasets and how the are produced (provenance records). For reproducibility they can simply re-generate all steps.
They don't include data version control, however, some integrate with community-specific research data portals which support versioned data publications (coarse-grained).
Workflow systems are usually quite complex and have a considerable learning curve. However, if there is a de-facto standard in your field of science then this is certainly the solution of choice.

The presented solution is generic and adopts the widely familiar git know-how for an easier learning curve.

\subsection{Combined Versioning and Reproducibility}

Very few generic tools for this combination exist.
DataLad is "a Python-based tool for the joint management of code, data, and their relationship, built on top of a versatile system for data logistics (git-annex) and the most popular distributed version control system (Git)." \cite{halchenko2021datalad} developed as Open Source software since 2013.
It provides the \shell{datalad} command with sub-commands very much like git.
Since it works on top of \shell{git} and \shell{git-annex}, all those sub-commands can be used directly, so can the integrated \shell{git} functionality in many IDEs and GUI tools.
As far as \shell{git} / \shell{git-annex} functionality is concerned, DataLad only adds convenience features and helpful information --- especially when DataLad really uses a collection of \shell{git} repositories and submodules under the hood.
For DataLad's added reproducibility functionality see Section~\ref{sect:datalad-reproducibility}.
DataLad is commonplace in several scientific communities and actively
developed and maintained by a consortium of US and German universities and research institutes.
There is demand for an HPC integration and HPC application cases have been published \cite{wagner2022fairly}.


Data Version Control (DVC) is an Open Source tool for managing and versioning of large datasets and processing pipelines in data-driven projects. It is advertised as "git for data" as well%
\footnote{The DVC webpage \url{https://dvc.org/doc/start}}
and is widely adopted for machine learning \cite{barrak2021co}. DVC started in 2017.
It uses its own command \shell{dvc} with a list of sub-commands. It keeps all metadata about large/binary files and md5 hashes in "metafiles". Only those are versioned with git (or other SCMs) under the hood. Large/binary data referenced by the metafiles can be stored locally or in external (cloud) storages.
The \shell{dvc stage add} command can be used to define steps of processing "pipelines" together with inputs and outputs. Then, the pipelines can be initially executed and later reproduced with \shell{dvc repro}. This works recursively and only re-executes on changed inputs.
DVC has no HPC integration and there are no publications combining DVC with HPC 
to the best of our knowledge. There seems to be little demand due to DVC's ML focus.

The rest of the paper is based on DataLad as the older project with a clearly documented demand for HPC \cite{wagner2022fairly}.

\subsection{The SLURM batch scheduler}

SLURM, short for Simple Linux Utility for Resource Management, is an open-source, highly scalable cluster management and job scheduling system widely used in HPC \cite{slurm}.
The three primary functions are (1) resource allocation, (2) job execution, and (3) queue management.
This paper focuses on SLURM  because it is the dominating job scheduler for HPC clusters worldwide.
With respect to the implementation there are specifics such as command names and data formats, still, the presented solution should serve well as a template for corresponding extensions for other job schedulers.

\section{Background: Machine-Actionable Reproducibility with DataLad}
\label{sect:datalad-reproducibility}

DataLad offers novel functionality for machine-actionable reproducibility beyond the \shell{git} and \shell{git-annex} functionality. This is available through the extra sub-commands \shell{datalad run} and \shell{rerun}.
Instead of running any command (from a simple script to full-blown simulation software) and then committing the resulting data to the repository \shell{datalad} \shell{run} should be used. Besides the actual command, it accepts the set of input and output files or directories as well as a commit message as parameters. Then it will perform the following steps:

\begin{enumerate}

\item Make sure that all specified input files which are annexed files are available, if not run \shell{datalad get}. Also, it unlocks%
\footnote{"Lock" and "unlock" are \shell{git-annex} terminology, basically ro and rw.}
the specified output files, which allows them to be modified and later re-committed.

\item Execute the given command, wait for completion, and evaluate the return code.

\item Afterwards, commit the output files to the data repository and add a structured JSON reproducibility record to the commit message, see \figurename{}~\ref{abb:datalad-run-commit-message} for an example.

\end{enumerate}


\begin{figure}[b]
\begin{scriptsize}
\begin{verbatim}
commit ceab44790a98ed28c4f159ed85228c28e437460d
Author: Andreas Knüpfer <a.knuepfer@hzdr.de>
Date: Mon Feb 3 07:18:05 2025 +0100

    [DATALAD RUNCMD] Solve N=14 with ...
    === Do not change lines below ===
    {
     "chain": [],
     "cmd": "./scripts/run.sh 14 more-arguments-here",
     "dsid": "d5f31a22-4f48-4f83-a9ff-093b1ff3bbda",
     "exit": 0,
     "extra_inputs": [],
     "inputs": [
      "data/halos/14/generate_14.data.csv.xz"
     ],
     "outputs": [
      "data/results/14/worker/report.json",
      "data/results/14/worker/result.csv.xz"
     ],
     "pwd": "."
    }
    ^^^ Do not change lines above ^^^
\end{verbatim}
\end{scriptsize}
\caption{Example reproducibility record in the \shell{git} log from \shell{datalad run}.}
\label{abb:datalad-run-commit-message}
\end{figure}

\noindent
Later, \shell{datalad rerun} will re-execute it from the current state of the repository. The preparatory steps are:

\begin{enumerate}
\setcounter{enumi}{3}
\item Clone a DataLad repository from gitlab or similar. 
\item Required binary executables need to be available, for example from the HPC environment (modules, Spack, etc.) or compiled from sources first. Scripts should ideally be part of the same repository or a submodule.
\end{enumerate}


\noindent
The git commit hash to be reproduced can be identified from the git log, e.g. with \shell{git blame}. Then \shell{datalad rerun} \shell{<commit-hash>} re-executes it.
Internally, DataLad will:

\begin{enumerate}
\setcounter{enumi}{5}

\item Fetch annexed files if not already present according to the \shell{"inputs"} in the reproducibility record. The current versions in the data repository will be used.

\item Execute the \shell{"cmd"} from the reproducibility record in the original directory \shell{"pwd"}.

\item Check if the \shell{"outputs"} files changed.
Commit altered outputs to the data repository. If nothing changed, no new commit is done.

\end{enumerate}


The typical use case is re-creating someone else's computational results from their DataLad repository.
Either to obtain the result data (because downloading is too expensive or restricted) or to check if the documented processing steps are sufficient. DataLad can even confirm that all results are bit-wise reproduced based on the file hashes \textit{without} actually downloading the original annexed files from the repository.

Among others, this would be most convenient for a paper review process. A DataLad repository can be shared as the reproducibility appendix. The simulation source code is in a sub-repository (git submodule) and the commit hashes to be rerun are specified. Then \shell{datalad rerun} would reproduce the paper's results in a fully automatic way. The reviewers could judge which results should be bitwise identical (for example a histogram from given values) or have minimal difference according to a suitable metric (for example results from iterative solvers) or just similar enough (for example run-time measurements as part of the paper's evaluation).

\subsection{Shortcomings of DataLad in HPC Environments}
\label{sect:shortcomings}

DataLad and the underlying \shell{git} and \shell{git-annex} commands
work in HPC environments in principle.
However, there are two critical reasons why DataLad is incompatible with HPC and SLURM.

\paragraph{Critical Shortcomings}

The fundamental conflict is that \shell{git} repositories expect no more than one \shell{git} command to be active at a time. Concurrent \shell{git} commands produce race conditions.
While \shell{git} implements some internal locking, it is \textbf{not safe} to run concurrent \shell{git} commands in the same repository. 
Yet, SLURM is supposed to run many jobs concurrently. Thus DataLad or git commands inside a SLURM job script would be dangerous.

The only alternative is to restrict to to one job \textit{scheduled} at any given time per repository. This could either mean serialization of SLURM jobs. Even though every job may be parallel inside, this would be a contradiction of HPC usage. 
Or it could be separate copies ("clones") per SLURM job. This is the best possible workaround
%
and the current state of the art when combining DataLad with HPC, see Wagner \textit{et al}. \cite{wagner2022fairly}. It places datalad commands
inside SLURM jobs which operate on separate clones of the same DataLad repository. 
This causes considerable redundancies and metadata stress for the parallel file system.


\paragraph{Critical Inefficiencies}
\label{sec:critical-inefficiencies}

In addition, there are critical inefficiencies if the existing DataLad reproducibility feature is used inside SLURM jobs.
\shell{Datalad} and the underlying \shell{git} / \shell{git annex} commands are essentially serial commands. Thus, there would be multiple serial sections inside a SLURM job which is considered inefficient in highly parallel jobs.

\paragraph{One step back in reproducibility in HPC?}

There is an additional disadvantages when using the workaround from \cite{wagner2022fairly} for both, DataLad itself and its aptitude for HPC.

The SLURM job script is useful for reproducibility since it defines which hardware/partition used, how many processes/threads employed, the memory requirements, runtime bounds, the software modules used, and the commands/job steps executed. This is certainly useful information for re-execution and it is machine-actionable. But since it contains a line \shell{datalad run <command> <arguments>} which needed to be changed to \shell{datalad rerun <commit-hash>} for re-execution, the SLURM script cannot be re-used without modification. Thus it is by definition no longer part of the machine-actionable reproducibility.

\section{Solution: The Datalad SLURM extension}

The presented solution is available from the Github repository \url{https://github.com/knuedd/datalad-slurm}%
\footnote{The branch \textit{runtime-evaluation} contains the runtime measurement data and the Jupyter notebooks to generate all diagrams of this paper.}
as Open Source Software and comes as a plug-in to any standard DataLad installation.
It works like follows to achieve the primary goals:
\begin{itemize}
    \item Allow many SLURM jobs to be scheduled and to run at the same time on the same DataLad repository.
    \item Avoid race conditions due to job input and output files.
    \item Generate reproducibility records in the git log and provide fully machine-actionable re-execution of jobs.
\end{itemize}

\subsection{Job execution}

Since DataLad commands must not be inside the SLURM job script, the solution uses two datalad invocations before and after the job. The first is \shell{datalad slurm-schedule} which is a prefix command to the the normal \shell{sbatch} call with the script filename and all arguments.
The job script doesn’t require any modification. It should be part of the data repository so that it is under version control itself.

\shell{datalad slurm-schedule} like \shell{datalad run} \textit{may} list required inputs for auto-retrieval of  annexed files and \textit{must} specify output files%
\footnote{Unlike \shell{datalad run} where specifying outputs is partly optional.}
or exclusive directories for them. \shell{datalad} \shell{slurm-schedule} checks that there are no conflicting outputs in the same repository for already scheduled jobs. If so it calls the \shell{sbatch} command to schedule the job. Otherwise it refuses to schedule the new job with a warning message to prevent a race condition.

The information about jobs \textit{"in flight"} (currently scheduled, not yet finalized) and their files is managed in a local SQLite database under \shell{./git/}. It is a single instance for all branches for the given clone of the DataLad repository.

\subsection{Job Finalization}

The DataLad call to finalize a job is \shell{datalad slurm-finish}. It has to be called some time after one or multiple SLURM jobs finished. If run with a SLURM job ID it will handle the specified job only, otherwise it will cover all jobs not handled before. The list of \textit{in flight} jobs is taken from the internal SQLite database. Jobs still pending or running are ignored. 
Canceled of failed jobs can either be discarded using the flag \shell{--close-failed-jobs} or handled like successful jobs using \shell{--commit-failed-jobs}. Once \shell{datalad} \shell{slurm-finish} successfully ran for a job its output files may be reused by future jobs; the would no longer cause conflicts and now there is a commit between the modifications.

\begin{figure}[b]
\begin{scriptsize}
\begin{verbatim}
commit e1c781afe46bfecd0e447fc0c239b1e0386da0cc
Author: Andreas Knüpfer <a.knuepfer@hzdr.de>
Date:   Fri Mar 14 11:39:40 2025 +0100

    [DATALAD SLURM RUN] Slurm job 11452054: Completed

    === Do not change lines below ===
    {
     "chain": [],
     "cmd": "sbatch slurm.sh",
     "dsid": "4928ddbc-d6fe-4fa4-bff7-25ec6a2dca88",
     "extra_inputs": [],
     "inputs": [],
     "outputs": [
      "test_01_output_dir_18",
      "log.slurm-11452054.out",
      "slurm-job-11452054.env.json"
     ],
     "pwd": "test_01_output_dir_18",
     "slurm_job_id": 11452054,
     "slurm_outputs": [
      "log.slurm-11452054.out",
      "slurm-job-11452054.env.json"
     ]
    }
    ^^^ Do not change lines above ^^^
\end{verbatim}
\end{scriptsize}
\caption{Example of a SLURM job reproducibility record in the \shell{git} log. Compare the counterpart from \shell{datalad run} in \figurename~\ref{abb:datalad-run-commit-message}.}
\label{abb:datalad-slurm-schedule-commit-message}
\end{figure}

The output files of the job will be added and their changes committed to the DataLad repository. 
Two extra outputs will be handled in the same way: the SLURM job output file and an additional file \shell{slurm-job-<id>.env.json} which contains the SLURM metadata from the job execution in JSON format. Then the JSON reproducibility record is added to the git log as part of the commit message, see also \figurename~\ref{abb:datalad-slurm-schedule-commit-message}.

\begin{figure}[h]
\centering
\includegraphics[scale=0.3]{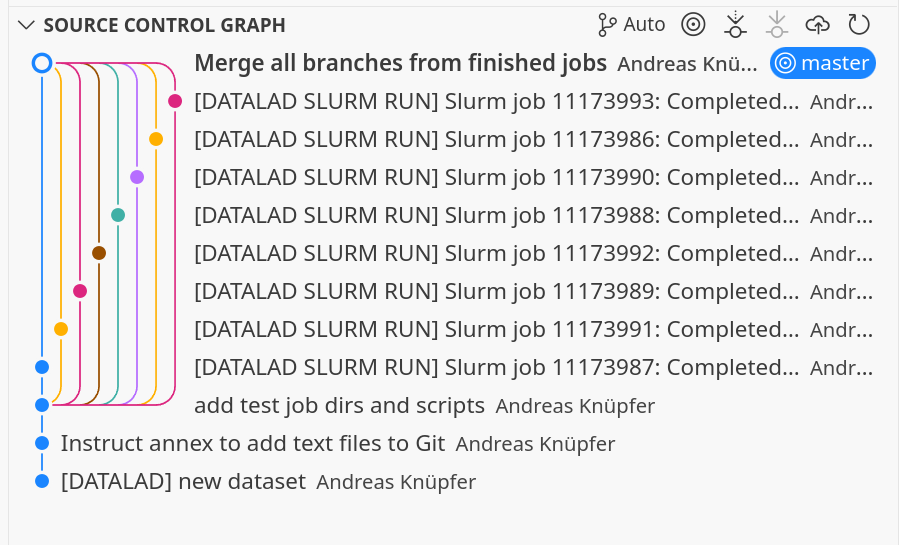}
\caption{Visualization of the \shell{git} commit graph when using the \shell{--octopus} switch for \shell{datalad slurm-finish}.}
\label{fig:git-commit-graph}
\end{figure}

\subsection{Re-Execution}

To reproduce or re-execute a SLURM job, \shell{datalad slurm-reschedule <commit-hash>} can be used similar to \shell{datalad rerun}. The commit needs to contain the \texttt{[DATALD SLURM RUN]} reproducibility record, compare \figurename~\ref{abb:datalad-slurm-schedule-commit-message} and all details for the job are taken from there. If the executables or scripts%
\footnote{For the SLURM job script also the newest version from the current branch is used. This allow to modify the partition or queue specifications, for example, if the job should be re-executed on a different HPC cluster.}
or the input files changed since the original job execution, then the latest versions are used.

For such rescheduled jobs there needs to follow a call to \shell{datalad slurm-finish} just like for original jobs and it will do the same finalization steps.

\subsection{Job Branches and Octopus Merge}

When many jobs are submitted and later collected and committed during \shell{datalad slurm-finish}, then the files and subdirs of each job become one commit with one commit hash. The commits from concurrent jobs will not be in a strict logical order but in a arbitrary order. 
A commit to a separate branch per job better reflects the concurrent nature of the SLURM jobs.
\shell{datalad slurm-finish --branches} will do this automatically. 
With \shell{--octopus} the same branching is done followed by a so called \textit{octopus merge}, i.e. a git merge with multiple branches at once, see \figurename~\ref{fig:git-commit-graph}.

\subsection{Array jobs}

Array jobs are a special SLURM feature to create many similar jobs with one \shell{sbatch} call.
They will be similar but work on different tasks, of course. A typical example is different parameter sets in a large parameter study where each case is independent from all others.
During \shell{datalad}\shell{slurm-schedule} an array job is like a single job with potentially very many inputs and outputs, no special options are required.
During \shell{slurm-finish} DataLad waits for all individual jobs to be finished. It will also check the job statuses individually.
If all report as "COMPLETED" the array job as a whole is considered successfully finished. Otherwise, one of \shell{--close-failed-jobs} or \shell{--commit-failed-jobs} has to be used.
The reproducibility record is created for the entire array job, thus it can only be rerun as a whole for now.

\subsection{Alternative job directories}
\label{sect:altdir}

For SLURM jobs, all job inputs and outputs need to be accessible from all compute nodes of a cluster, therefore, they need to be on a parallel file system.
Unfortunately, DataLad or \shell{git} repositories may have unfavorable effects on parallel file systems, see Section~\ref{sec:eval:finish} below.
The \shell{--alt-dir <dir>} option of \shell{datalad slurm-schedule} allows the DataLad repository to stay on a local file system where \shell{git} is much faster. It will:

\begin{enumerate}

\item Construct the real working directory under the given alternative directory on a parallel file system with the same relative path as in the repository.

\item Deep copy all input files and directories for the job.

\item Submit the job from inside the alternative directory.

\end{enumerate}

\noindent
Later, \shell{datalad slurm-finish} will do the reverse 
\begin{enumerate}
\setcounter{enumi}{3}
\item Copy all output files back to the repository.
\end{enumerate}

\noindent
before its normal steps. Copying to/from the alternative directory is easy when all inputs and outputs are specified. All other aspects work like without the \shell{--alt-dir} option.
See Section~\ref{sec:eval:finish} for the advantages of this flavor.

\section{Performance Evaluation}
\label{sec:eval}

The evaluation will focus on the runtime overhead compared to normal SLURM jobs as the "cost" of DataLad's reproducibility benefits on HPC. Only \shell{slurm-schedule} and \shell{slurm-finish} are benchmarked because \shell{slurm-reschedule} is the same as \shell{slurm-schedule} only with all parameters taken from the git history.
The experiments used DataLad version 1.1.5 and the DataLad SLURM plugin version 0.1.1.

\subsection{Experiment setup}

Each benchmark series uses a newly created Datalad repository for a number of jobs with one exclusive directory per job in a nested hierarchy.
Each job consists of the job script without any input files. It produces a small text file and a small binary file (becoming an annexed file) as output. The SLURM output file \shell{log*.out} and the metadata file \shell{slurm-job-*.env.json} are implicit outputs, thus, each job creates 4 outputs. Two more variants of this produce 4 or 8 additional files per job; one half text files, one half binary files.
The test repositories are located on the GPFS parallel file system.

For each of the three cases with 4, 8, or 12 outputs per job there is an counterpart where the DataLad repository is located on a local XFS file system. The \shell{--alt-dir} option is used to run those SLURM jobs, see Section~\ref{sect:altdir}.
Finally, the same kind of jobs are scheduled with pure SLURM \shell{sbatch} as a comparison case.

\subsection{Job Submission Overhead}

During job submission overhead may arise due to:

\begin{enumerate}
\item Loading the DataLad Python package (esp. the first time)
\item Checking the state of the data repository
\item Retrieving necessary input files if they are annexed files
\item Check for conflicts of output arguments
\item Calling SLURM under the hood (sbatch, sacct, ...)
\item Copying input files if \shell{--alt-dir} is used
\end{enumerate}

For individual jobs submitted manually all of them are uncritical, only (1) and (2) may be annoying if they take several seconds. In all cases, (3) is considered neutral, because without DataLad those files would need to be copied in another way.
The overheads from (2), (4), and (5) are relevant for automated submission for many jobs.

\begin{figure*}[t]
\centering
\includegraphics[scale=0.44]{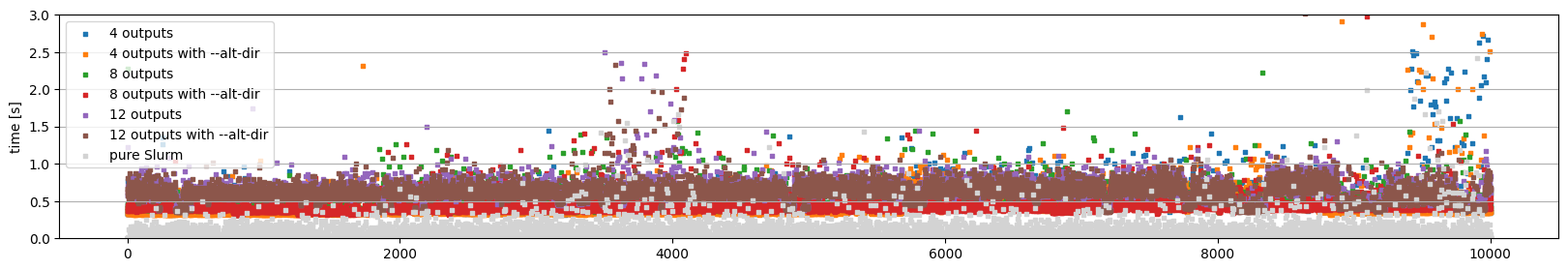}
\includegraphics[scale=0.44]{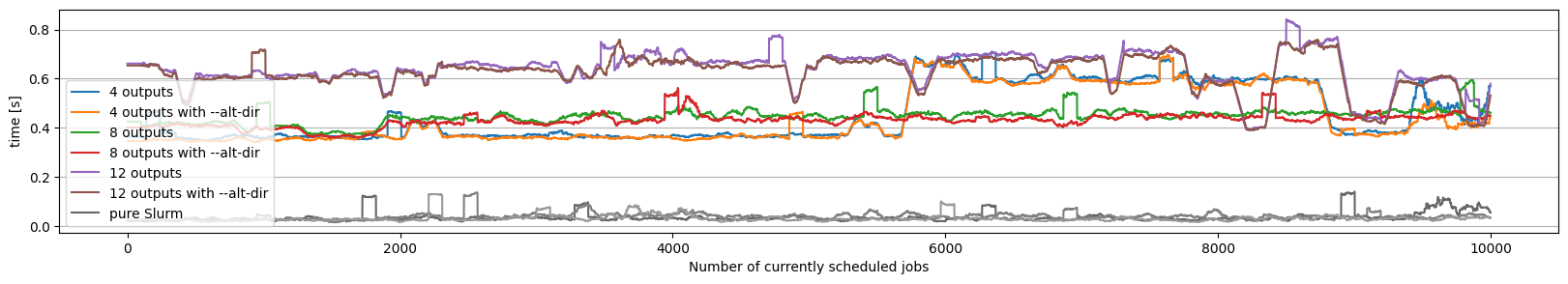}
\caption{Runtime for the \shell{datalad slurm-schedule} command over the number of currently scheduled jobs (in colors) compared to SLURM sbatch calls (shown in gray). Top: scatter plot showing noisy data. Bottom: The same data as rolling averages which even out outliers but maintain the sum.}
\label{fig:schedule-time-rollingavg}
\end{figure*}

\begin{figure*}[t]
\centering
\includegraphics[scale=0.44]{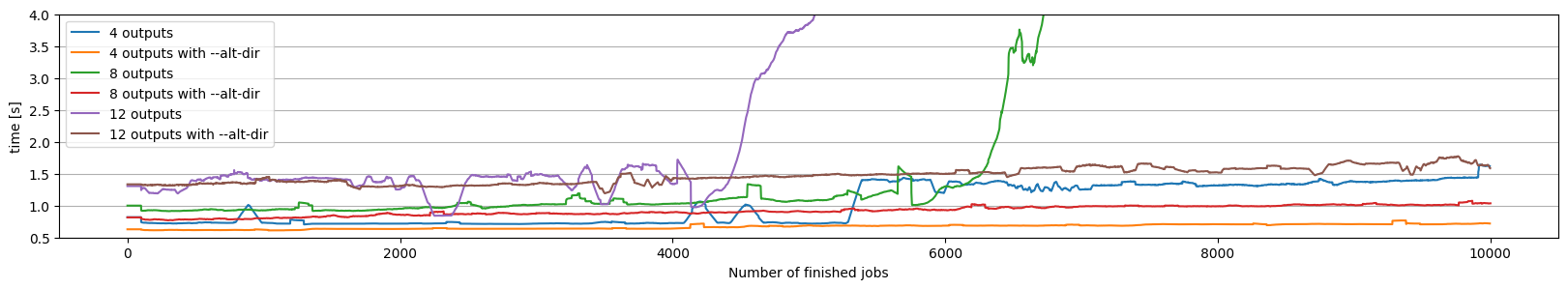}
\caption{Runtime per \shell{datalad slurm-finish} call over the number of jobs already committed to the repository. It is shown as rolling averages over a window of 100 values. The green curve continues to grow up to well over $10$s, the purple one well over $17$s.}
\label{fig:finish-time-rollingavg_extra}
\end{figure*}

\paragraph{Observations}

\begin{figure}[h]
\vspace{5mm}
\centering
\includegraphics[scale=0.44]{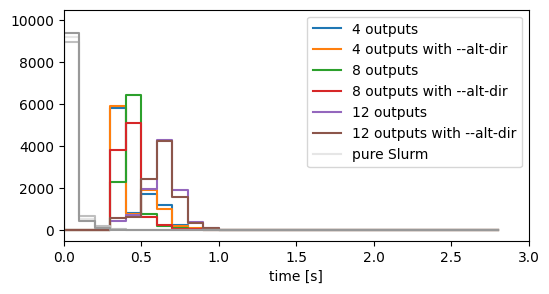}
\caption{The histogram of the runtimes from \figurename~\ref{fig:schedule-time-rollingavg}. It shows that the six different DataLad schedule cases with varying numbers of outputs per job have an average overhead of $0.3$s to $0.7$s over the "pure SLURM" cases (gray). All cases show a "long tail" of very few much slower cases.}
\label{fig:schedule-time-histogram}
\end{figure}

\figurename~\ref{fig:schedule-time-rollingavg} shows the runtimes when submitting $10\,000$ job from a repository in varying configurations. \figurename~\ref{fig:schedule-time-histogram} gives the histograms for the same values.
For all cases the measurements are very noisy
%
%
with most values well below $1$s but outliers over $10$s.
There are phases with more outliers for all cases -- the jobs of different types have been scheduled in a round robin fashion, so job $n$ from each group experienced roughly the same noise level or SLURM background load.

In \figurename~\ref{fig:schedule-time-rollingavg} (bottom) the same runtimes as rolling averages over a window of 100 jobs. This evens out the outliers and makes the general behavior much clearer without changing the sum of the runtimes.
Most pure SLURM cases have a runtime below $0.05$s, the DataLad jobs are mostly between $0.35$s and $0.7$s while all cases show phases with higher plateaus.
The DataLad cases with the same number of outputs behave very similar, regardless if \shell{--alt-dir} is used or not.

\paragraph{Interpretation}

The noisy behavior is shared by all cases and is most likely caused by step (5), i.e. SLURM. 
There is no notable difference between a repository on the parallel file system or on the local file system.
The overheads from steps (1), (2), (4), and (6) produce a roughly constant extra runtime of $0.35s$ to $0.6s$ compared to pure SLURM. Step (3) does not apply here.

There is a trend that more output arguments lead to a longer runtime, yet this is less pronounced than the plateau phases.
There is no indication that the number of jobs or the number of files already in the repository or the size of the internal SQLite job database have a consistent effect.

All in all, this runtime behavior is uncritical and suitable for very many jobs per DataLad repository. Here it was demonstrated well beyond the number of jobs where one should switch to array jobs.

\subsection{Job Finalization Overhead}
\label{sec:eval:finish}

During finalization via \shell{datalad slurm-finish} the following steps are performed which might cause overheads:

\begin{enumerate}

\setcounter{enumi}{6}

\item Copy back output files to the repository if \shell{--alt-dir} was used during scheduling.

\item Commit outputs to the repository, add the JSON reproducibility record to the commit message.

\item Convert binary files (or other configured file types) to annexed files by \shell{git-annex}.

\end{enumerate}

The experiments continue with the DataLad repositories mentioned above but not the pure SLURM counterparts, as they don't have a finishing step. After all submitted jobs have been finished they will be handled individually by \shell{datalad slurm-finish --slurm-job-id <id>} to measure their individual runtimes.

\paragraph{Observations}

\begin{figure}[h]
\vspace{7mm}
\centering
\includegraphics[scale=0.44]{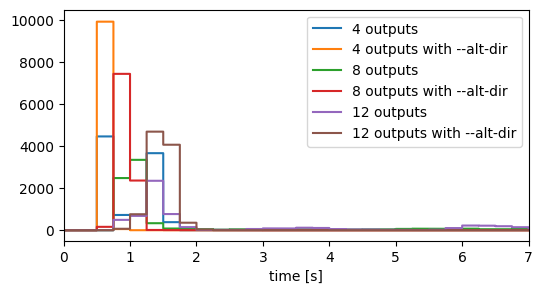}
\caption{Histogram of the runtimes from \figurename~\ref{fig:finish-time-rollingavg_extra}. There is a clear trend that the runtimes grow with more outputs files per job, regardless if \shell{--alt-dir} was used or not. The long tail for all series is cut at $7s$. There are no pure SLURM counterparts here.}
\label{fig:finish-time-histogram}
\end{figure}

\figurename~\ref{fig:finish-time-rollingavg_extra} shows the runtime of the \shell{datalad slurm-} \shell{finish} calls as rolling averages and \figurename~\ref{fig:finish-time-histogram} shows the histogram.

Two of the three cases without \shell{--alt-dir} behave in an undesirable manner with a strong growth after a certain point well beyond $10$s (green) or well beyond $17$s (purple) per job on average.
All three cases using \shell{--alt-dir} show very moderate growth with the number of jobs.

\paragraph{Interpretation}

The strong increase of the green and purple curves is caused by steps (8) and (9). Step (7) is not involved in those cases.
The issue is caused by the number of files in a DataLad repository when it is placed on a parallel file system. It starts at around $50\,000$ files in a repository ($>4000$ jobs with $12$ outputs or $>6000$ jobs with $8$ outputs).
If one comes back to one of those repositories later, then the same behavior persists.

When repositories with the same number of files are kept on a local file system, then there is only a very moderate increase in the runtime over the number of jobs committed. This increase is more pronounced with more outputs per job. Still, this is tolerable and uncritical even with this quite large number of individual jobs.

The underlying reason for this behavior is the file system caching, most likely the directory entry ("dentry") and inode caching. Once there are more directories or files than the cache can cover, the cache misses require access to the storage which is slower. On a local SSD or NVMe file system the slow down is very moderate and almost negligible. On a parallel cluster file system which is a remote file system, the slowdown is dominated by the network latency and is thus quite substantial. Further benchmarks are underway to test this hypothesis.

The suggested solution using the \shell{--alt-dir} approach works well with small files as shown here. The copy-back step (7) does not show any adverse effect. It is expected to show some influence with large files, though.
For much larger files, there will be a break-even point when copying is slower than the delay due to a repositories on a parallel file system.

All in all, \shell{datalad slurm-finish} is slower than \shell{slurm-schedule} and more sensitive to the underlying file system. The decision when to use \shell{--alt-dir} should depend on the behavior of \shell{datalad} \shell{slurm-finish}.

Even when the DataLad repositories are located on a parallel file system there is a long phase when \shell{datalad slurm-finish} behaves nicely. This may be enough when there are fewer but much larger files.
If the delay starts to show one can switch to the \shell{--alt-dir} fashion before the next batch of jobs is to be scheduled. There is no need to decide  during the creation of a DataLad repository. 

More benchmarks are planned to study the effect closer on different clusters, different parallel file systems (Lustre, Weka, and others), and with varying file sizes.

\section{Summary and Outlook}


The paper presents an extension to DataLad which enables data versioning and machine-actionable reproducibility for HPC and the SLURM batch scheduling system. It shows the incompatibilities of the existing approach with HPC, discusses the proposed solution, and evaluates the implementation.
The final solution is suitable for large numbers of jobs and large data repositories. It costs one to two seconds time overhead per job altogether which is slower than scheduling pure SLURM jobs but acceptable for the provided benefits. It includes an option to prevent undue timing or scaling behavior with very many jobs resp. very many files on parallel file systems.


We see three main directions of future work.
Firstly, the behavior of git on parallel file systems has to be investigated further. It will be interesting to compare different parallel file systems 
and to study how the number of files vs. the file sizes influence the slowdown of git operations.

Secondly, the \shell{--alt-dir} functionality should be compared to the \shell{git worktree}%
\footnote{\url{https://git-scm.com/docs/git-worktree}} feature. A \textit{linked worktree} is a separate copy of a branch in an external path. If this behaves well on a parallel file system, then this should become another way to keep the main repository in a local file system while SLURM jobs can still access their required files on a parallel file system. This will be less selective but may be more efficient when almost all files will be required for jobs. 

Thirdly, the \shell{datalad} \shell{slurm-reschedule} functionality for array jobs should be extended so that individual jobs can be re-scheduled. This would be useful, if most jobs in an array job ran successfully but a few failed.

\section*{Acknowledgments}

We'd like to thank the HPC operations team at HZDR for their support and their tolerance with our extensive SLURM tests.
Furthermore, we are grateful for the helpful feedback from the DataLad community.

\section*{Declaration on Generative AI}

The ScienceOS.ai AI assistant (\url{https://app.scienceos.ai/}) was used for research about related work and helped to find additional relevant papers to cite. Also it was used to draft the related work paragraphs, however, they have been significantly edited by the authors afterwards.

\bibliographystyle{IEEEtran}
\bibliography{IEEEabrv,bibtex.bib}

\begin{thebibliography}{10}
\providecommand{\url}[1]{#1}
\csname url@samestyle\endcsname
\providecommand{\newblock}{\relax}
\providecommand{\bibinfo}[2]{#2}
\providecommand{\BIBentrySTDinterwordspacing}{\spaceskip=0pt\relax}
\providecommand{\BIBentryALTinterwordstretchfactor}{4}
\providecommand{\BIBentryALTinterwordspacing}{\spaceskip=\fontdimen2\font plus
\BIBentryALTinterwordstretchfactor\fontdimen3\font minus
  \fontdimen4\font\relax}
\providecommand{\BIBforeignlanguage}[2]{{%
\expandafter\ifx\csname l@#1\endcsname\relax
\typeout{** WARNING: IEEEtran.bst: No hyphenation pattern has been}%
\typeout{** loaded for the language `#1'. Using the pattern for}%
\typeout{** the default language instead.}%
\else
\language=\csname l@#1\endcsname
\fi
#2}}
\providecommand{\BIBdecl}{\relax}
\BIBdecl

\bibitem{mala1}
\BIBentryALTinterwordspacing
A.~Cangi, L.~Fiedler, B.~Brzoza, K.~Shah, T.~J. Callow, D.~Kotik, S.~Schmerler,
  M.~C. Barry, J.~M. Goff, A.~Rohskopf, D.~J. Vogel, N.~Modine, A.~P. Thompson,
  and S.~Rajamanickam, ``Materials learning algorithms (mala): Scalable machine
  learning for electronic structure calculations in large-scale atomistic
  simulations,'' 2024. [Online]. Available:
  \url{https://arxiv.org/abs/2411.19617}
\BIBentrySTDinterwordspacing

\bibitem{mala2}
\BIBentryALTinterwordspacing
L.~Fiedler, N.~Hoffmann, P.~Mohammed, G.~A. Popoola, T.~Yovell, V.~Oles,
  J.~Austin~Ellis, S.~Rajamanickam, and A.~Cangi, ``Training-free
  hyperparameter optimization of neural networks for electronic structures in
  matter,'' \emph{Machine Learning: Science and Technology}, vol.~3, no.~4, p.
  045008, 10 2022. [Online]. Available:
  \url{https://dx.doi.org/10.1088/2632-2153/ac9956}
\BIBentrySTDinterwordspacing

\bibitem{wilkinson2016fair}
M.~D. Wilkinson, M.~Dumontier, I.~J. Aalbersberg, G.~Appleton, M.~Axton,
  A.~Baak, N.~Blomberg, J.-W. Boiten, L.~B. da~Silva~Santos, P.~E. Bourne
  \emph{et~al.}, ``The fair guiding principles for scientific data management
  and stewardship,'' \emph{Scientific data}, vol.~3, no.~1, pp. 1--9, 2016.

\bibitem{klump_2020_3772870}
\BIBentryALTinterwordspacing
J.~Klump, L.~Wyborn, M.~Wu, R.~Downs, A.~Asmi, G.~Ryder, and J.~Martin,
  ``Principles and best practices in data versioning for all datasets big and
  small,'' Apr. 2020. [Online]. Available:
  \url{https://doi.org/10.15497/RDA00042}
\BIBentrySTDinterwordspacing

\bibitem{eval_of_rdm_architectures}
B.~Heinrichs, M.~Politze, and M.~A. Yazdi, ``Evaluation of architectures for
  fair data management in a research data management use case,'' in
  \emph{Proceedings of the 11th International Conference on Data Science,
  Technology and Applications - DATA}, INSTICC.\hskip 1em plus 0.5em minus
  0.4em\relax SciTePress, 2022, pp. 476--483.

\bibitem{RDM_version_control}
A.~Freund, H.~Hajiabadi, and A.~Koziolek, ``Exploring existing tools for
  managing different types of research data,'' in \emph{INFORMATIK 2024}.\hskip
  1em plus 0.5em minus 0.4em\relax Bonn: Gesellschaft für Informatik e.V.,
  2024, pp. 2181--2193.

\bibitem{huber2020aiida}
S.~P. Huber, S.~Zoupanos, M.~Uhrin, L.~Talirz, L.~Kahle, R.~H{\"a}uselmann,
  D.~Gresch, T.~M{\"u}ller, A.~V. Yakutovich, C.~W. Andersen \emph{et~al.},
  ``Aiida 1.0, a scalable computational infrastructure for automated
  reproducible workflows and data provenance,'' \emph{Scientific data}, vol.~7,
  no.~1, p. 300, 2020.

\bibitem{uhrin2021aiida}
\BIBentryALTinterwordspacing
M.~Uhrin, S.~P. Huber, J.~Yu, N.~Marzari, and G.~Pizzi, ``Workflows in aiida:
  Engineering a high-throughput, event-based engine for robust and modular
  computational workflows,'' \emph{Computational Materials Science}, vol. 187,
  p. 110086, 2021. [Online]. Available:
  \url{https://www.sciencedirect.com/science/article/pii/S0927025620305772}
\BIBentrySTDinterwordspacing

\bibitem{galaxy2024galaxy}
``The galaxy platform for accessible, reproducible, and collaborative data
  analyses: 2024 update,'' \emph{Nucleic acids research}, vol.~52, no.~W1, pp.
  W83--W94, 2024.

\bibitem{gruning2018galaxyreproducibility}
B.~Gr{\"u}ning, J.~Chilton, J.~K{\"o}ster, R.~Dale, N.~Soranzo, M.~Van
  Den~Beek, J.~Goecks, R.~Backofen, A.~Nekrutenko, and J.~Taylor, ``Practical
  computational reproducibility in the life sciences,'' \emph{Cell systems},
  vol.~6, no.~6, pp. 631--635, 2018.

\bibitem{halchenko2021datalad}
Y.~O. Halchenko, K.~Meyer, B.~Poldrack, D.~S. Solanky, A.~S. Wagner, J.~Gors,
  D.~MacFarlane, D.~Pustina, V.~Sochat, S.~S. Ghosh \emph{et~al.}, ``Datalad:
  distributed system for joint management of code, data, and their
  relationship,'' \emph{Journal of Open Source Software}, vol.~6, no.~63, p.
  3262, 2021.

\bibitem{wagner2022fairly}
A.~S. Wagner, L.~K. Waite, M.~Wierzba, F.~Hoffstaedter, A.~Q. Waite,
  B.~Poldrack, S.~B. Eickhoff, and M.~Hanke, ``Fairly big: A framework for
  computationally reproducible processing of large-scale data,''
  \emph{Scientific data}, vol.~9, no.~1, p.~80, 2022.

\bibitem{barrak2021co}
A.~Barrak, E.~E. Eghan, and B.~Adams, ``On the co-evolution of ml pipelines and
  source code-empirical study of dvc projects,'' in \emph{2021 IEEE
  International Conference on Software Analysis, Evolution and Reengineering
  (SANER)}.\hskip 1em plus 0.5em minus 0.4em\relax IEEE, 2021, pp. 422--433.

\bibitem{slurm}
A.~B. Yoo, M.~A. Jette, and M.~Grondona, ``Slurm: Simple linux utility for
  resource management,'' in \emph{Job Scheduling Strategies for Parallel
  Processing}, D.~Feitelson, L.~Rudolph, and U.~Schwiegelshohn, Eds.\hskip 1em
  plus 0.5em minus 0.4em\relax Berlin, Heidelberg: Springer Berlin Heidelberg,
  2003, pp. 44--60.

\end{thebibliography}

\end{document}